LA-UR-12-24491

Title: Diode magnetic-field influence on radiographic spot size

Author(s): Ekdahl, Carl A. Jr.

Intended for: Report

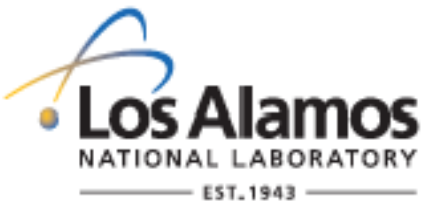

# Diode magnetic-field influence on radiographic spot size

*Carl Ekdahl*

## I. INTRODUCTION

Flash radiography of hydrodynamic experiments driven by high explosives is a well-known diagnostic technique in use at many laboratories [1,2,3]. The Dual-Axis Radiography for Hydrodynamic Testing (DARHT) facility at Los Alamos was developed for flash radiography of large hydrodynamic experiments [4,5,6]. Two linear induction accelerators (LIAs) produce the bremsstrahlung radiographic source spots for orthogonal views of each experiment ("hydrotest"). The 2-kA, 20-MeV Axis-I LIA creates a single 60-ns radiography pulse [4,7]. For time resolution of the hydrotest dynamics, the 1.7-kA, 16.5-MeV Axis-II LIA creates up to four radiography pulses by slicing them out of a longer pulse that has a 1.6-µs flattop [5,6,8,9,10,11]. Both axes now routinely produce radiographic source spot sizes having full-width at half-maximum (FWHM) less than 1 mm.

To further improve on the radiographic resolution, one must consider the major factors influencing the spot size:

- Beam convergence at the final focus
- Beam emittance
- Beam canonical angular momentum
- Beam-motion blur
- Beam-target interactions

Beam emittance growth and motion in the accelerators have been addressed by careful tuning [10,12]. Defocusing by beam-target interactions has been minimized through tuning of the final focus solenoid for optimum convergence and other means [7,11]. Finally, the beam canonical angular momentum is minimized by using a "shielded source" of electrons. An ideal shielded source creates the beam in a region where the axial magnetic field is zero, thus the canonical momentum zero, since the beam is born with no mechanical angular momentum. It then follows from Busch's conservation theorem that the canonical angular momentum is minimized at the target, at least in principal.

In the DARHT accelerators, the axial magnetic field at the cathode is minmized by using a "bucking coil" solenoid with reverse polarity to cancel out whatever solenoidal beam transport field exists there. This is imperfect in practice, because of radial variation of the total field across the cathode surface, solenoid misalignments, and long-term variability of solenoid fields for given currents. Therefore, it is useful to quantify the relative importance of canonical momentum in determining the focal spot, and to establish a systematic methodology for tuning the bucking coils for minimum spot size. That is the purpose of this article.

Section II provides a theoretical foundation for understanding the relative importance of the canonical momentum. Section III describes the results of simulations used to quantify beam parameters, including the momentum, for each of the accelerators. Section IV compares the two accelerators, especially with respect to mis-tuned bucking coils. Finally, Section IV concludes with a methodology for optimizing the bucking coil settings.

## II. THEORY

External magnetic fields at the electron source (cathode) in the DARHT accelerators can affect the radiographic spot size at the final focus. For example, an electron beam born without mechanical angular momentum (rotation) in an axial (solenoidal) magnetic field acquires rotation when it leaves the field in order to conserve canonical angular momentum (the sum of mechanical and field momentum). If the final focus is in a field free region, the outward centrifugal "force" due to such rotation prevents the beam from focusing as tightly as it otherwise would. Therefore, radiographic accelerators incorporate means for eliminating the magnetic field at the cathode (shielded source), or more practically, nulling the magnetic flux linking the cathode, which is directly proportional to the field momentum. A beam born with no field momentum will produce the smallest spot in a field-free region.

As usual, the beam envelope equation [13,14,15] provides an excellent theoretical framework for understanding the beam physics. Here, we use it to investigate the relative importance of the factors influencing the spot size, including canonical angular momentum. To first order, the beam envelope equation for an azimuthally symmetric beam coasting downstream from the final focus solenoid to the target can be written as

$$\frac{d^2 a}{dz^2} = -k_\beta^2 a + \frac{K}{a} + \frac{\varepsilon^2 + (P_\theta / \beta \gamma m_e c)^2}{a^3} \tag{1}$$

Here, $a(z)$ is the radius of an equivalent uniform beam, defined by $a = \sqrt{2}\, r_{rms}$, where $r_{rms}$ is the root-mean-square radius of the actual beam profile.

The first term in Eq. (1) is solenoidal focusing, with $k_\beta^2 = (2\pi B_z / \mu_0 I_A)^2$, where the Alfven limiting current is

$$I_A = \frac{4\pi}{\mu_0} \frac{m_e c}{e} \beta \gamma = 17.08 \beta \gamma \text{ kA} \tag{2}$$

Here, $\beta = v_e / c$, and $\gamma = \sqrt{1 - 1/\beta^2}$.

The second term is due to space charge and can be either defocusing (as in vacuum) or focusing (if the beam space-charge is neutralized by ions). The generalized perveance is

$$K = (2 I_b / \beta^2 I_A) \{(1 - f_e) - \beta^2 (1 - f_m)\} \tag{3}$$

In this equation, the space-charge neutralization factor is $f_e = n_i / n_e$ for a beam with uniform beam density $n_e$ propagating through an ion background having with uniform density $n_i$. The current neutralization factor $f_m = I_r / I_b$, where $I_r$ is the induced return current, is usually unimportant, so the perveance can be simplified to $K = \left(2I_b / \beta^2 I_A\right)\{(1-f_e) - \beta^2\}$, or $K = \left(2I_b / \beta^2 I_A\right)\{1/\gamma^2 - f_e\}$. This shows that the term is focusing when ion neutralization is large enough; $f_e > 1/\gamma^2$. For an un-neutralized beam in vacuum the term is defocusing with $K = \left(2/\beta^2\gamma^2\right)(\nu/\gamma)$, where $\nu/\gamma = I_b/I_A$ is the Budker parameter. Thus, ions produced by beam-target interaction can have a profound effect on spot size; e.g., as ion density increases, the focal point can move upstream, causing the radiographic spot to enlarge with time [7].

The final term in Eq. (1) is always defocusing. In this term $\varepsilon$ is the beam emittance, given by

$$\varepsilon = 2\sqrt{\langle r^2 \rangle \left[\langle r'^2 \rangle + \langle (v_\theta/\beta c)^2 \rangle\right] - \langle rr' \rangle^2 - \langle rv_\theta/\beta c \rangle^2} \qquad (4)$$

where brackets denote ensemble averages (moments). Also in the final term, the canonical angular momentum is $P_\theta = \gamma m_e a^2 \omega - eaA_\theta$, where $A_\theta$ is the vector potential of the solenoidal field, and $\omega$ is the rotational frequency of the beam. By Busch's theorem, $P_\theta$ is conserved [14], so it is equal to the initial value at the cathode $P_\theta = P_{\theta K} = -er_K A_\theta(r_K)$ where, $r_K$ is the radius of the cathode. By integrating the $z$ component of $\mathbf{B} = \nabla \times \mathbf{A}$ one finds

$$r_K A_{\theta K} = \int_0^{r_K} rB_z(r)\,dr = \Phi_m / 2\pi \qquad (5)$$

where the integration is over the face of the cathode, and $A_{\theta K} = A_\theta(r_K)$. Thus, $r_K A_{\theta K}$ and $P_\theta$ are proportional to the magnetic flux linking the cathode, $\Phi_m$, rather than $B_z$ on axis, unless $B_z$ is constant in radius, which is impossible in the fringe fields of solenoids.

Equation (1) can be easily solved for $da/dz$ by reduction to quadrature and integration. Then, setting $da/dz = 0$ yields a transcendental expression for the beam size at a waist. Although the waist size is not necessarily the smallest focal spot at the target [16], for DARHT parameters it is close enough to ignore the error. For typical DARHT parameters at the focus, and $f_e = 0$, the size of the waist is dominated by the emittance, and is approximately $a_{\min} \approx \varepsilon_{eff} / a'_0$, where the effective emittance $\varepsilon_{eff}$ is the quadrature sum of the emittance and canonical angular momentum, and $a'_0$ is the beam convergence after the final focus solenoid. Thus, in the absence of beam-target interactions, and for given $a'_0$ (established by initial beam size and focal length), the size of the spot is completely determined by the emittance and angular momentum.

If the source is perfectly shielded, $P_{\theta K} = 0$, and the spot size is minimized. If it is not perfectly shielded, the field angular momentum adds in quadrature with the emittance, and thereby increases the spot size. Therefore, it is reasonable to compare the initial value of angular momentum to the beam emittance measured with methods such as focusing magnet scans [8] in order to determine the impact of imperfect shielding on the focal spot. As expressed in Eq. (1), the momentum can be directly compared with the normalized emittance, $\varepsilon_n = \beta\gamma\varepsilon$, which would be conserved through the accelerators in the absence of non-linear forces [14].

Thus, one can consider a total effective normalized emittance,

$$\varepsilon_{eff\,n}^2 = \varepsilon_n^2 + (P_\theta/m_e c)^2 = \varepsilon_n^2 + (er_K A_{\theta K}/m_e c)^2 \qquad (6)$$

and an equivalent normalized emittance resulting from the magnetic fields at the cathode can be defined by

$$\varepsilon_{eqn} \equiv (P_\theta/m_e c) = (er_K A_{\theta K}/m_e c) \qquad (7)$$

which is numerically equal to $5.866 \times 10^8 r_K A_{\theta K}$ π-mm-mr when the units of $r_K A_{\theta K}$ are T-m$^2$.

As an illustration of the sensitivity of the effective emittance to errors in the bucking coil setting consider that an error of 1 Gauss (10$^{-4}$ T) over 200 cm$^2$ (approximately the area of the Axis-II cathode) produces an equivalent normalized field emittance of 186 π-mm-mr, which is about the same as the emittance $\varepsilon_n$ from a perfectly shielded cathode predicted by PIC and gun-design code simulations. The component of the earth field normal to the DARHT-II cathode is ~0.24 G, so it alone can add ~50 π-mm-mr to the total emittance. For comparison, the total emittance measured downstream of the DARHT accelerators is of order 1,000 π-mm-mr.

## III. SIMULATIONS

The magnetic field angular momentum at the cathode was found from detailed 2-D simulations of the DARHT diode magnetic field. These simulations were initially performed in conjunction with simulations of diode performance for both axes[17,18]. The applied magnetic fields were simulated using finite-element methods based on a conformal triangular mesh model of the DARHT diodes used for finite-element simulations [19]. The applied magnetic field was calculated with the TriComp/PerMag code [20], using accurate models of the solenoids in the diode and accelerator cells, including magnetic materials. One of the powerful features of PerMag is that it has numerical and graphical output of $rA_\theta$, so parametric studies of equivalent emittance are easily accomplished.

### *A. DARHT Axis-I Simulations*

The DARHT Axis-I diode uses a velvet "explosive-emission" cold cathode as a source of electrons for the space-charge limited beam current. The standard 5.08-cm (2-inch) diameter cathode produces a 1.75-kA beam accelerated to 3.8 MeV in a 60-ns pulse [21]. Unless otherwise noted, Axis-I simulations reported here used a 5.08-cm cathode diameter, with the cathode recessed to the

The ~2-m diameter Axis-I bucking coil is mounted externally to the diode vacuum tank. The magnetic field in the DARHT Axis-I diode was obtained by modeling the bucking coil and anode solenoid as ideal sheet solenoids having the dimensions and locations as modeled in the XTR envelope code used to tune the accelerator. The base-case field was simulated for 100 A energizing the anode solenoid, and -14.51 A energizing the bucking coil. During operations, and retuning the accelerator, the ratio of these two currents, $\left|I_{buck}/I_{anode}\right| \equiv k_{buck} = 0.1451$, has been kept constant. Therefore, no matter what the operational tune, $P_\theta$ is proportional to the anode solenoid field (and energizing current).

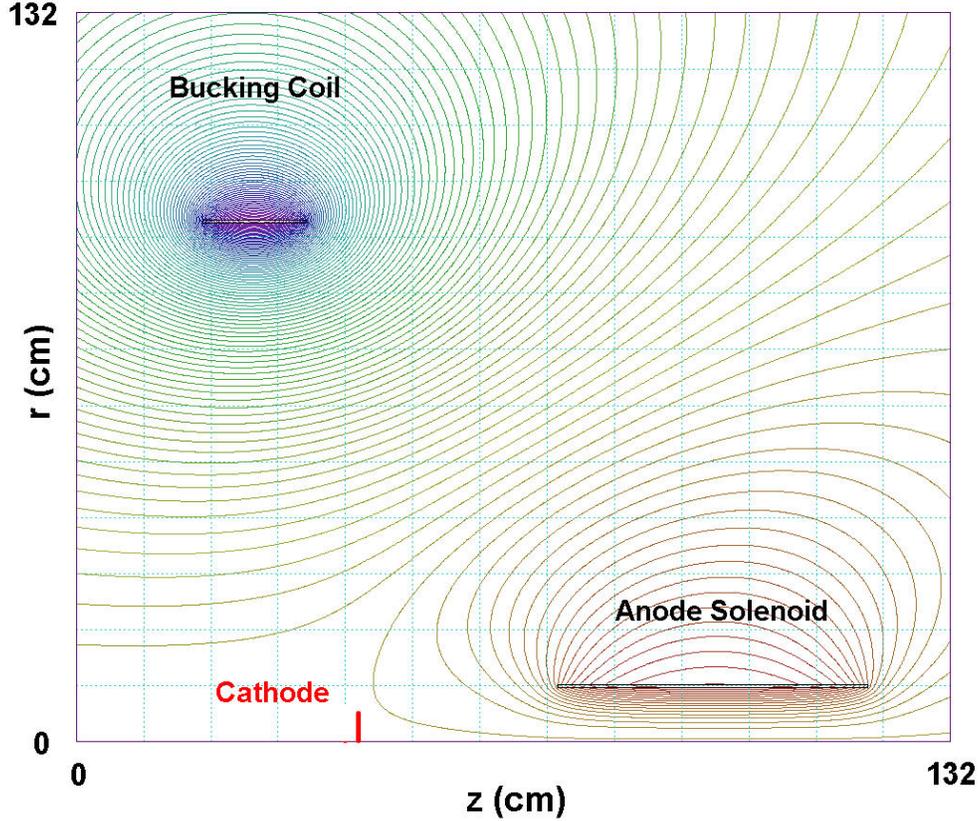

Figure 1: Contours of constant $rA_\theta$ in the DARHT Axis-I diode region. These contours are proportional to the magnetic flux, $\Phi_m = 2\pi rA_\theta$. Also shown are the locations of the anode solenoid, bucking coil, and cathode.

Figure 1 shows the locations of the anode solenoid, bucking coil, and cathode. Also shown in Fig. 1 are contours of constant $rA_\theta$, which is proportional to the magnetic flux, $\Phi_m = 2\pi rA_\theta$, and also to the equivalent emittance. The large diameter bucking coil external to the diode provides a relatively large volume of equivalent emittance that is significantly smaller than the ~1,000 π-mm-mr normalized emittance measured experimentally [21].

The standard 5.08-cm (2-inch) diameter Axis-I cathode is recessed by ~3-mm below the flat shroud surface to produce a 1.75-kA beam[18]. This depth may be slightly varied for the desired current and beam profile, so it is useful to know the sensitivity of $P_\theta$ with depth. Figure 2 shows $rA_\theta$ at the radius of the cathode as a function of axial position, and the value of the equivalent normalized emittance. Here, the sign of $\varepsilon_{eqn}$ indicates the rotational direction of the angular momentum, and has no significance, because $\varepsilon_{eqn}$ is added to $\varepsilon_n$ in quadrature. Clearly, the recessed cathode is close to optimum. The equivalent emittance at the cathode is only ~0.5 π-mm-mr with the nominal 100-A setting for the anode magnet, and the sensitivity to the exact cathode depth is only ~1.1 π-mm-mr/mm. These are indeed insignificant compared to the ~1,000 π-mm-mr measured emittance, and there is little to be gained by tinkering with the present magnet settings.

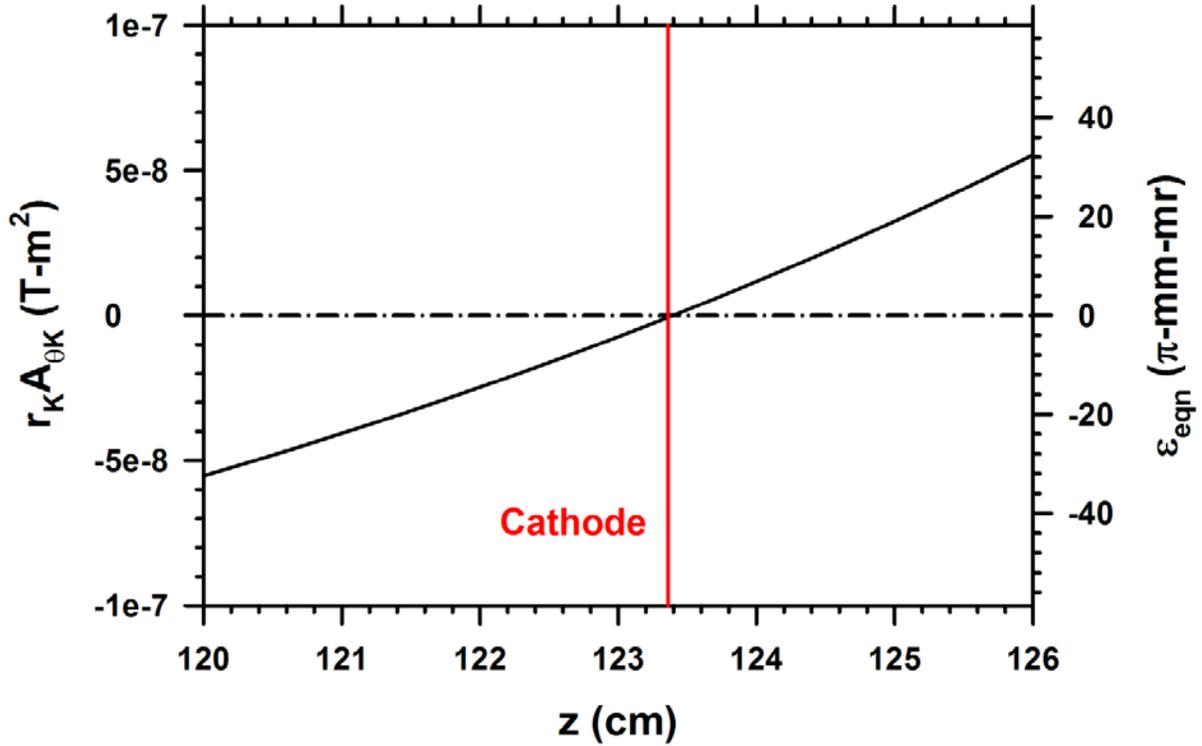

Figure 2: Variation of $rA_\theta$ with axial position at the Axis-I cathode radius. The right scale is the equivalent normalized emittance corresponding to the flux linking the cathode. The nominal location of the recessed cathode is indicated by the vertical solid red line.

During operations the ratio of exciting currents for the anode solenoid and bucking coil is fixed $|I_{buck}/I_{anode}| \equiv k_{buck} = 0.1451$. It follows that the sensitivity of equivalent emittance is linearly related to either. For example, with this fixed ratio, the sensitivity to the anode solenoid current is ~0.5 π-mm-mr/100 A, and 0.0345π-mm-mr/A for the bucking coil. On the other hand, it would be interesting to know the sensitivity of the equivalent emittance to variations in bucking-coil current with arbitrary $k_{buck}$. Figure 3 illustrates this variation, which has a sensitivity of 10.4 π-mm-mr/A near the optimum.

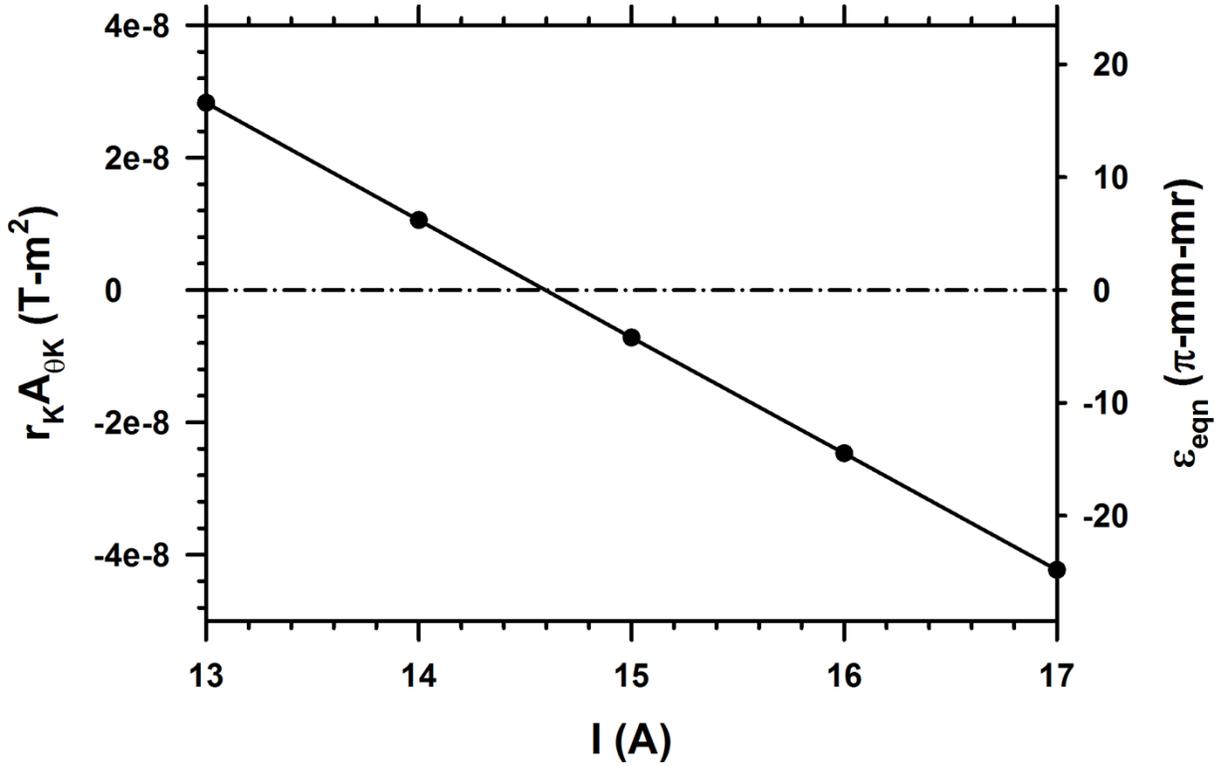

Figure 3: DARHT Axis-I $rA_\theta$ and equivalent emittance as functions of the bucking coil current. For these data the anode solenoid current was 100 A.

It is worth noting that this well-shielded configuration is not achieved by zeroing the axial field on axis at the cathode. It is obvious that when the field varies with radius, as in the Axis-1 cathode region, the total flux will be nulled only if the field at the center is opposite to the field at the edge of the cathode. This is illustrated in Fig. 4, which shows $B_z(r)$ across the recessed cathode face when the cathode is recessed 2.69 mm, where the flux linking the cathode would be exactly nulled by the bucking coil.

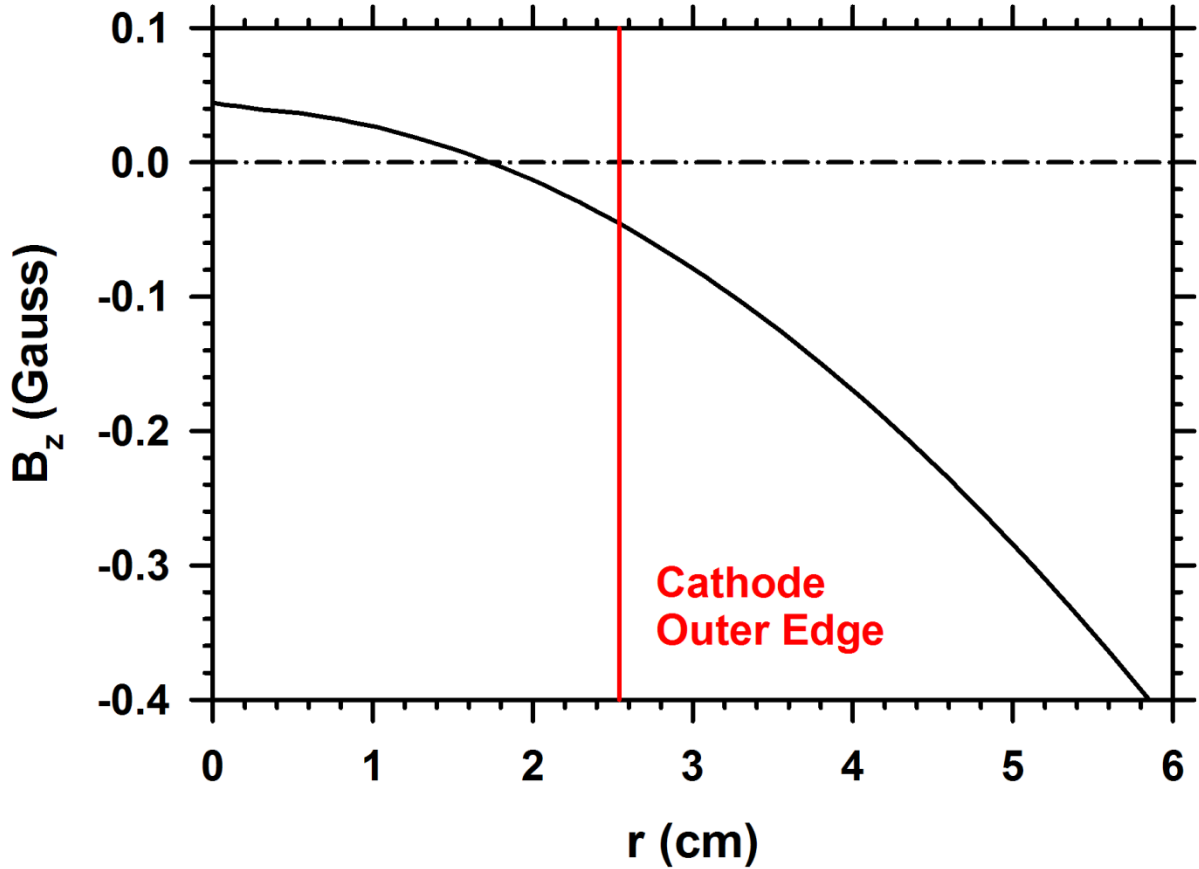

Figure 4: Axial magnetic field across the face of an Axis-I 2-inch diameter cathode recessed by 2.69 mm. This position is where the total flux linkage (and equivalent emittance) is zero.

## B. DARHT Axis-II Simulations

The source of electrons for the DARHT Axis-II beam is a 16.5-cm diameter hot dispenser cathode. As presently operated, this produces a 1.68-kA space-charge limited beam accelerated to 2.2 MeV in the diode [17].

In contrast to Axis-I, the Axis-II bucking coil is only ~ 0.4-m diameter, and is mounted in the vacuum, behind the cathode in the high voltage dome. Moreover, the ratio of current in the bucking coil to that in the anode magnets is not fixed, as it is on Axis-I. The solenoids included in the magnetic field model are shown in Fig. 5, which also shows the location of the cathode, and contours of constant $rA_\theta$. Locations of components of the diode were determined from the engineering ProE model, along with physical measurements of the anode-cathode shroud separations [23]. Figure 6 is a blow up of the region near the cathode surface.

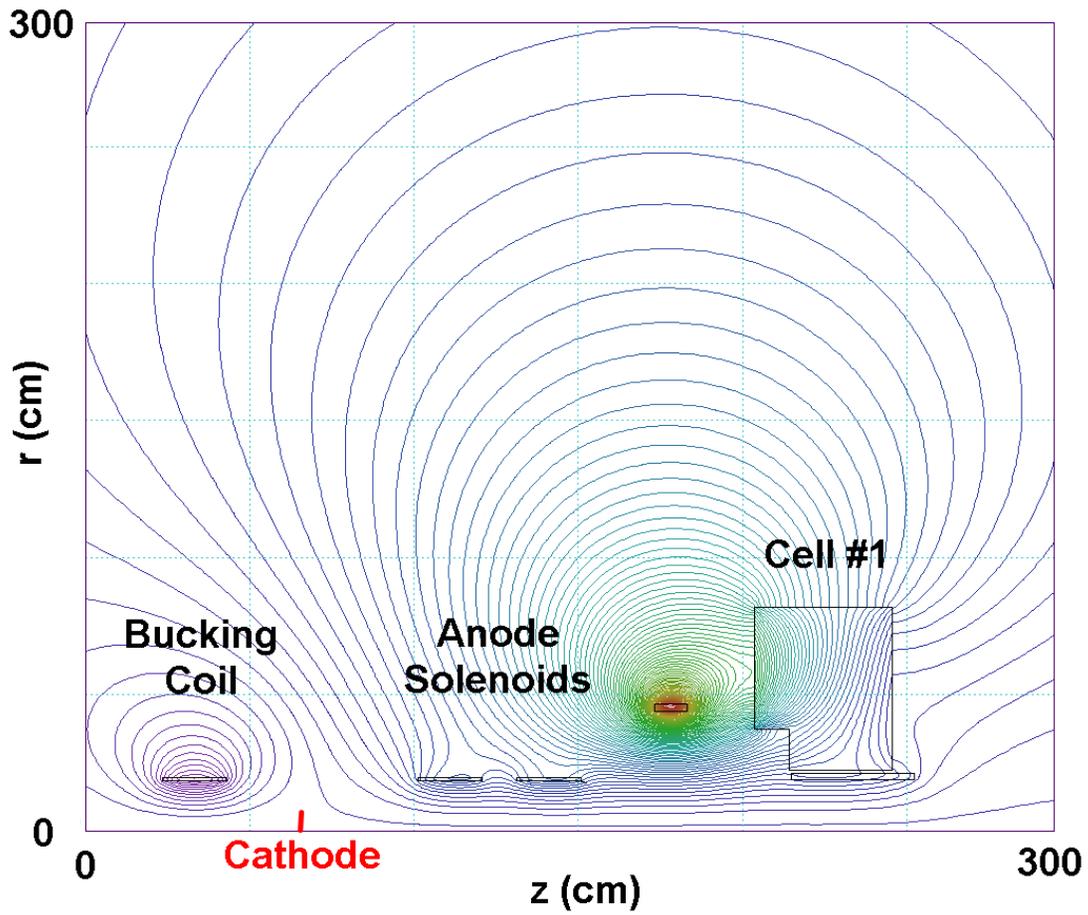

Figure 5: Contours of constant $rA_\theta$ in the DARHT Axis-II diode region. These contours are proportional to the magnetic flux, $\Phi_m = 2\pi rA_\theta$. Also shown are the locations of the anode solenoids, first acceleration cell, bucking coil, and cathode.

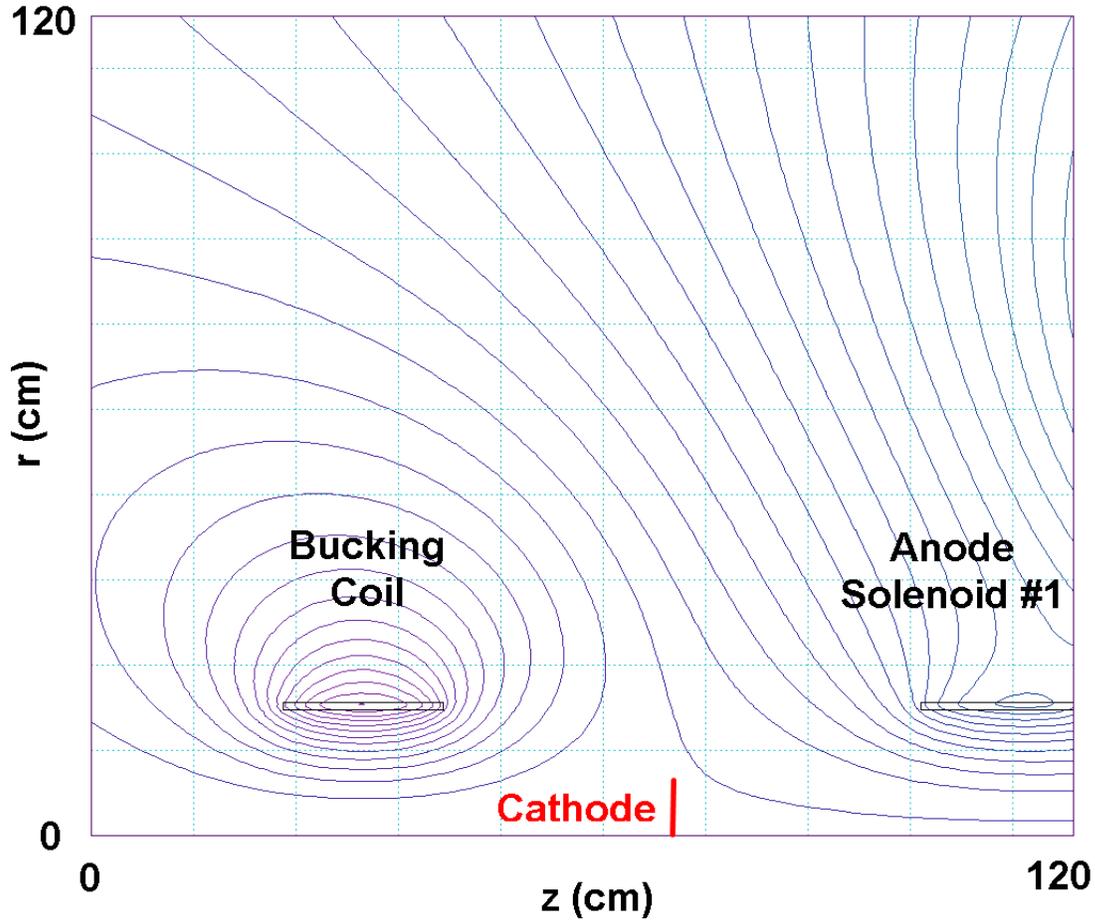

Figure 6: Contours of constant $rA_\theta$ in the DARHT Axis-II diode region. These contours are proportional to the magnetic flux, $\Phi_m = 2\pi rA_\theta$. Also shown are the locations of the anode solenoid, bucking coil, and cathode.

Figure 7 shows the variation of $rA_\theta$ with axial position at the cathode radius for the 17 A drive current used since May, 2011. Obviously, this setting of the bucking coil in this simulation is far from optimum; the equivalent emittance is 164 π-mm-mr, which is almost equal to the mechanical emittance predicted for the beam from LSP and TRAK simulations[12]. Figure 8 shows $rA_\theta$ and the equivalent emittance $\varepsilon_{eqn}$ as a function of current to the bucking coil. According to these simulation results, the flux linking the cathode would be nulled with a bucking coil current setting of ~17.5 A. The sensitivity to the solenoid current is large, ~277 π-mm-mr/A. This is ~27 times the Axis-I sensitivity.

It is also evident by comparing Fig. 7 to Fig. 2 that the sensitivity of emittance to cathode position is about 100 times that of Axis 1. The large Axis-II sensitivity to cathode location (~50 π-mm-mr/mm) is the result of the large cathode size and the small bucking coil size. This makes accurate knowledge of diode component locations essential for reliable simulations. For this reason alone, it is advisable to determine the correct bucking coil setting from experimental data. Indeed, the present setting was determined based on Hall-effect probe measurements of the axial field at the location of the cathode surface, but those may not have been accurate enough because of the large sensitivity to dimensional variance (as indicated by Fig. 7). Therefore, it is advisable to re-determine the proper bucking coil setting based on direct measurements of the minimum spot size (an option not available when the setpoint was initially determined).

Figure 9 shows the axial magnetic field at the cathode surface for three different values of current in the bucking coil. Clearly, 17.5 A is a better solution for zero total flux linking the cathode. Compared with Axis-I (Fig. 3), the Axis-II geometry produces an opposite radial field gradient at the cathode. Moreover the total variation across the cathode is almost twice that of the variation across the Axis-I cathode, again emphasizing the more difficult task of nulling the equivalent emittance on Axis-II.

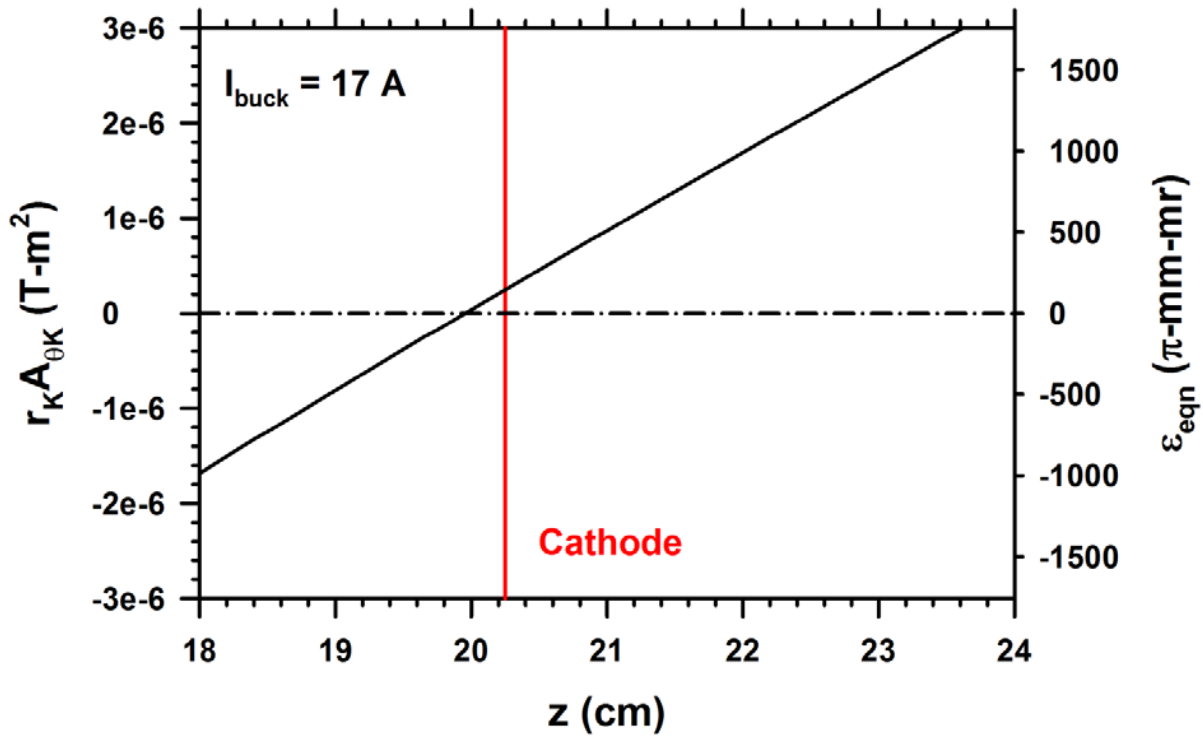

Figure 7: Variation of $rA_\theta$ with axial position at the Axis-II cathode radius. The right scale is the equivalent normalized emittance corresponding to the flux linking the cathode. The location of the cathode is indicated by the vertical solid red line.

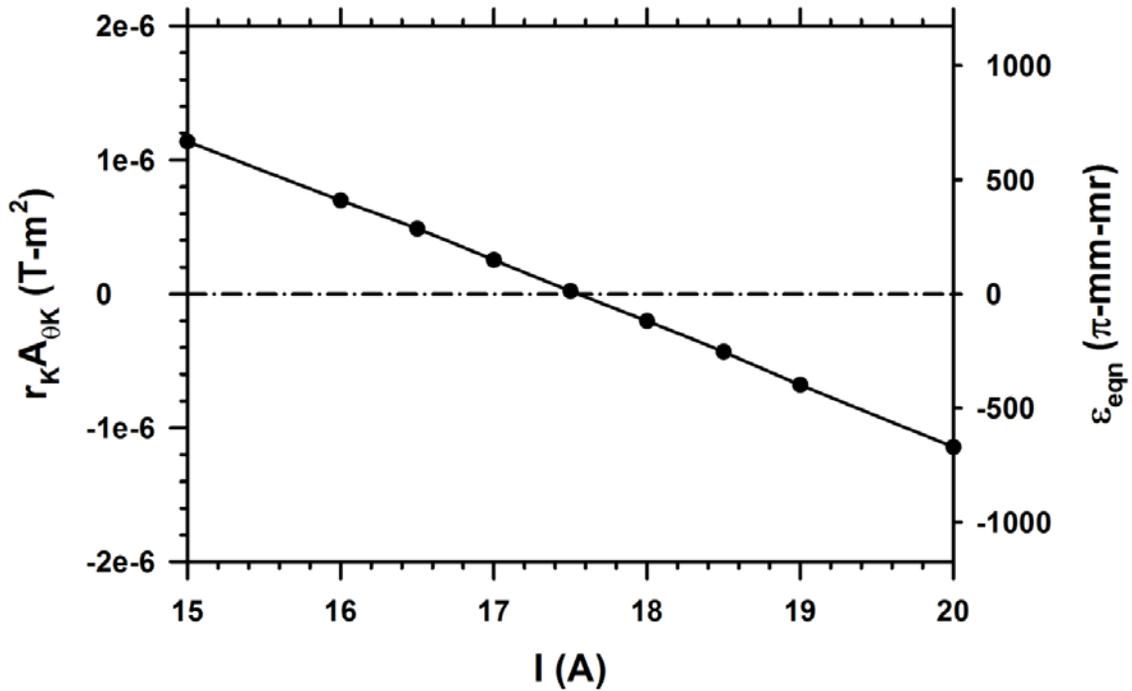

Figure 8: DARHT Axis-II $rA_\theta$ and equivalent emittance as functions of the bucking coil current.

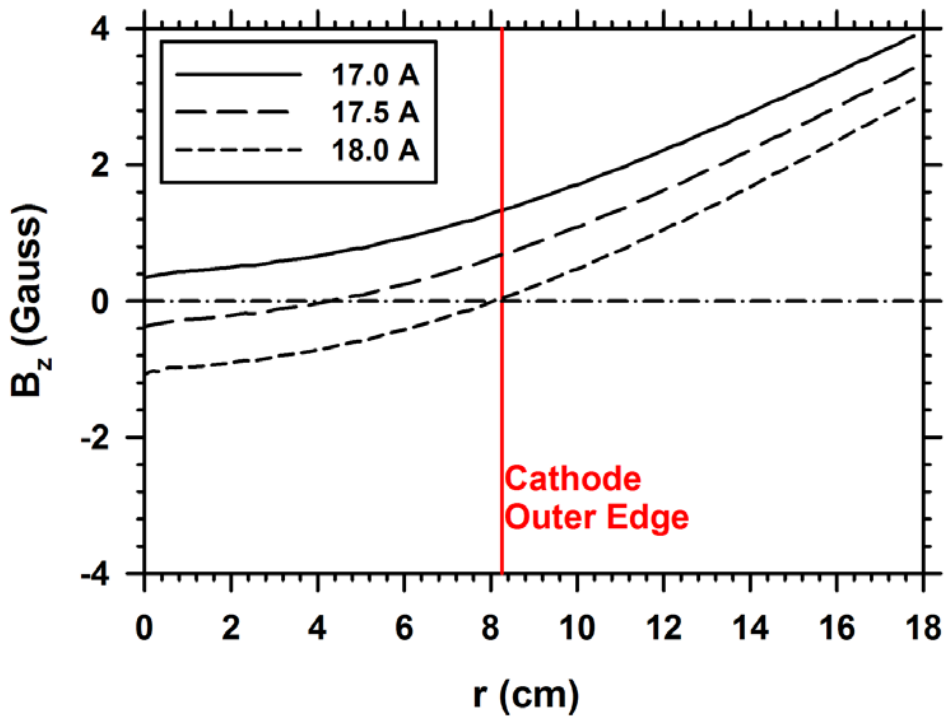

Figure 9: Axial magnetic field across the face of an Axis-II cathode for three different currents in the bucking coil. The 17.5 Ampere simulation produces almost no total linked flux (or equivalent emittance).

## IV. DISCUSSION

The major difference in field geometry between Axis-I and Axis-II is the aspect ratio, defined as $AR = \ell_{ba} R_{anode} / R_{buck}^2$, where $R_{buck}$ and $R_{anode}$ are the bucking coil and anode solenoid radii, and $\ell_{ba}$ is the center-to-center spacing. A large aspect ratio is associated with a long, skinny geometry, and a small aspect ratio is associated with a short, fat geometry. Inspection of Fig. 1 and Fig. 5 clearly shows this difference, which is quantified by comparing the aspect ratios (see Table I). From these simulations it is apparent that low aspect ratio leads to a reduced sensitivity to dimensional tolerances and magnet current variations, and is thus the preferred geometry whenever possible.

The other significant difference between the two axes is the physical size of the cathode. The larger size of the Axis-II cathode makes it much more sensitive to field variations.

The net result of these differences is that the unwanted equivalent emittance produced in the diode, and presumably the resulting enlargement of the spot size, is much more difficult to eliminate on Axis-II.

Table I: Comparison of Axis-I and Axis-II dimensions and simulation results

| Parameter | Symbol | Units | Axis-I | Axis-II |
|---|---|---|---|---|
| Bucking coil radius | $R_{buck}$ | cm | 94.2 | 19.05 |
| Anode solenoid radius | $R_{anode}$ | cm | 9.91 | 19.05 |
| Magnet Ratio | $R_{buck} / R_{anode}$ | | 9.5 | 1 |
| Spacing | $\ell_{ba}$ | cm | 122.9 | 79.1 |
| Cathode Diameter | $D_K$ | inch | 2 | 6.5 |
| Aspect Ratio | $AR$ | | 0.137 | 4.0 |
| Cathode Radius | $R_K$ | cm | 2.54 | 8.255 |
| Field Variation | $\Delta B_z$ | G | 0.54 | 1.05 |
| Current Sensitivity | $d\varepsilon_{eqn} / dI_{buck}$ | π-mm-mr/A | 10.4 | 277 |
| Cathode Position Sensitivity | $d\varepsilon_{eqn} / dz_K$ | π-mm-mr/mm | 1.1 | 50 |

The original bucking coil settings were based on Hall-probe field measurements at the cathode position. However, determination of the correct bucking coil current using a Hall probe is difficult because of the high sensitivity of the equivalent emittance to small errors in the measurements. Therefore, one should consider direct measurement of minimum spot size as a function of magnet current, a technique not available in the past. This can be accomplished by adjusting the final focus magnet for smallest spot size at each bucking coil current setting. (Changing the bucking coil setting slightly changes the tune, so the final focus will need to be re-adjusted for smallest spot.) The bucking coil currents are then varied until an absolute minimum is obtained. In following this procedure, one should use a single kicked pulse short enough that the spot size is not dominated by beam-target effects like ion defocusing [7,24,25,26].

# V. CONCLUSIONS

Because of its high aspect ratio magnet geometry and large area cathode, the DARHT Axis-II field angular momentum at the cathode is much more sensitive to dimensional inaccuracies and uncertainties in current settings than Axis-I. Since the equivalent emittance due to this effect could contribute to the radiographic spot size, it is important to determine the bucking coil current that minimizes flux linked by the cathode. This is best done experimentally using a single short pulse to minimize competing ion effects. The bucking coil current can be tuned for minimum spot size by adjusting the final focus magnet for smallest spot size at each of several bucking coil current settings. Once the proper current is established, its ratio to the anode solenoid currents should be fixed in the accelerator operating program (as in done on Axis-I).

# VI. ACKNOWLEDGEMENTS

The author acknowledges interesting discussions with B. Trent McCuistian, David C. Moir, and Martin E. Schulze on this and related beam physics topics. This research was performed under the auspices of the U. S. Department of Energy and the National Nuclear Security Administration under contract DE-AC52-06NA25396.